# Adaptive Sliding Mode Controller and Observer for Altitude and Attitude Control of a Quadrotor

K. Telli
Energetic System Modeling Laboratory
University of Biskra
Biskra, Algeria
telli.khaled@gmail.com

M. Boumehraz
Energetic System Modeling Laboratory
University of Biskra
Biskra, Algeria
m.boumehraz@univ-biskra.dz

A. Titaouine
Energetic System Modeling Laboratory
University of Biskra
Biskra, Algeria
a.titaouine@univ-biskra.dz

*Abstract*— **In this paper an adaptive sliding mode control approach for a quadrotor stabilization and trajectory tracking is presented. The closed loop control consists of three parts; the first part is quadrotor altitude and attitude stabilization, and trajectory tracking. Second part is used for parameters estimation where we focus in mass estimation, while the third part is the full states observation. Disturbances, sensors noise and parameter uncertainties, are taken into consideration. Sliding mode control law and observer adaptation are developed based on Lyapunov stability principle. Numerical simulations show the effectiveness of the proposed control technique**.

## I. INTRODUCTION

Nowadays, unmanned aerial vehicles (UAV) have known a growing interest due to their wide range of applications such as military, telecommunications, rescue operations, surveillance, monitoring, agriculture, delivery of goods and emergency medical intervention [1-3]. A quadcopter has several advantages compared to a conventional helicopter, like mechanical simplicity and high maneuverability. In addition, quadcopter provides a large lift thrust force that makes payload capacity increasing comparing to a helicopter. However, the major drawback of quadrotors is energy consumption due to the use of four actuators.

An UAV is required to perform risky or tedious missions under extreme operational conditions that are not feasible for conventional piloted flights [3].

Moreover, UAV's flight control systems are considered to be mission critical, as a flight control system failure could result in either the loss of the UAV or an unsuccessful mission. UAVs in operation remain remotely piloted, with autonomous flight control restricted to the primary modes of flight such as attitude hold, track hold along a straight track, from waypoint to waypoint, and a minimal level of control of flight or loiter maneuvers. While linear or gain-scheduled linear controllers may be typically adequate for these operations, it is often found that the UAV dynamics is nonlinear and that the linear controllers will not meet the performance requirements even when the gains are scheduled. This is due to the adverse coupling between the longitudinal and lateral dynamics that necessitates additional feedback. Furthermore, the uncertainty in the parameters of the mathematical model of the UAV and the environmental uncertainties are significant, and this can be another factor in making the design of long-endurance, high integrity autonomous flight controllers based on linear models that are extremely challenging.

Several advanced control approaches have been proposed to handle the nonlinear aerodynamics and kinematic effects, actuator saturations and rate limitations, modeling and parameters uncertainty to meet the increasing requirements on stability, performance and reliability.

Model predictive control [4], feedback linearization [5], backstepping control [6], robust control [7-10], fuzzy control [11], sliding mode control (SMC) [12-16] and various other techniques have been developed meet the increasing demands on the flight performance where the demand include attitude stabilization and tracking control [17-19].

SMC [20-22] is one of the most important control techniques to handle nonlinear systems with external disturbances. However, the main drawback of sliding mode control is the so-called chattering phenomenon which can excite undesirable high-frequency dynamics. Several methods have been proposed in order to reduce or eliminate the chattering phenomenon of SMC like the use of approximation function in place of the signum function or High order sliding mode control (HOSMC) which retains the property of robustness, ensure finite time convergence and also reduces the chattering.

Since the quadrotor mass is the dominated parameter, an adaptive sliding mode controller is proposed in order to stabilize the quadrotor and to compensate additive perturbations and parameter uncertainties related to the mass of the quadrotor. Adaptation law is developed based on a Lyapunov approach to assure asymptotic stability.

The rest of the paper is structured as follows. The quadrotor nonlinear dynamic model is presented in Section II, Section III describes the controller synthesis while the observer synthesis is presented in Section IV. The results of numerical simulation are shown in Section V. Finally, section VI concludes this paper.

## II. QUADRATIC DYNAMIC MODEL

A quadrotor is an UAV equipped with four rotors controlled independently as shown in Fig.1. The motion of the quadrotor results from changes in the speed of the rotors. The variation of the rotor speeds changes the lift forces and create motion. Thus, increasing or decreasing the four propellers speeds together generates vertical motion.

The pitch rotation is generated by varying the speeds of rotors 2 and 4. The roll rotation is adjusted by changing the speed of rotors 1 and 2. The yaw rotation results from the difference between speeds of the two pairs of propellers. All these operations should be performed while keeping the total thrust constant so that the altitude remains unchanged.

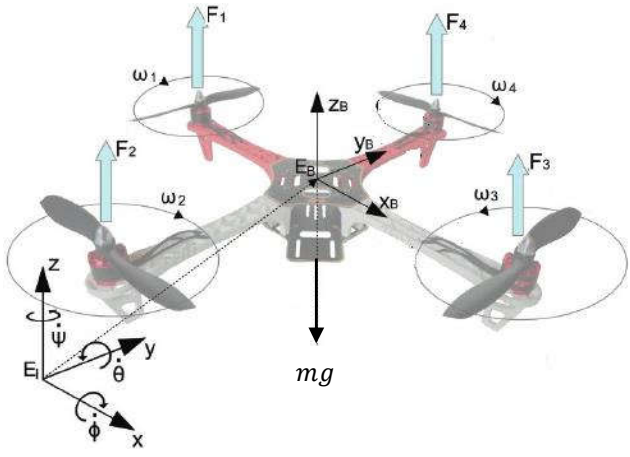

Fig. 1. Quadrotor configuration and applied forces

The quadrotor is a highly nonlinear, multivariable, strongly coupled, dynamically unstable and underactuated system. The quadrotor has six degrees of freedom with only four actuators.

Fig. 1. Shows the quadrotor systems and the force applied, where {B} is a body inertial frame fixed at quadrotor gravity center, and {E} is an earth fixed frame.

Let us consider the quadrotor structure and propellers rigid and symmetrical, the quadrotor dynamics can be described by the following equations [21]:

$$\ddot{x} = \frac{1}{m} [(\cos\phi \sin\theta \cos\psi + \sin\phi \sin\psi)U_1] \quad (1)$$

$$\ddot{y} = \frac{1}{m} [(\cos\phi \sin\theta \sin\psi - \sin\phi \cos\psi)U_1] \quad (2)$$

$$\ddot{z} = \frac{1}{m} (\cos\phi \cos\theta)U_1 - g \quad (3)$$

$$\ddot{\phi} = \frac{1}{I_{xx}} [(I_{yy} - I_{zz})\dot{\psi}\dot{\theta} - J_r \bar{\Omega}\dot{\theta} + lU_2] \quad (4)$$

$$\ddot{\theta} = \frac{1}{I_{yy}} [(I_{zz} - I_{xx})\dot{\psi}\dot{\phi} - J_r \bar{\Omega}\dot{\phi} + lU_3] \quad (5)$$

$$\ddot{\psi} = \frac{1}{I_{zz}} [(I_{xx} - I_{yy})\dot{\phi}\dot{\theta} - dU_4] \quad (6)$$

With:

$$\begin{pmatrix} U_1 \\ U_2 \\ U_3 \\ U_4 \end{pmatrix} = \begin{pmatrix} b & b & b & b \\ -b & 0 & b & 0 \\ 0 & -b & 0 & b \\ d & -d & d & -d \end{pmatrix} \begin{pmatrix} \omega_1^2 \\ \omega_2^2 \\ \omega_3^2 \\ \omega_4^2 \end{pmatrix} \quad (7)$$

And

$$\bar{\Omega} = \omega_1 - \omega_2 + \omega_3 - \omega_4 \quad (8)$$

where $\phi$, $\theta$, and $\psi$ angles represent the angler coordinates, $x$, $y$ and $z$ are the translational coordinates in space, $\omega_1$, $\omega_2$, $\omega_3$ and $\omega_4$ are the rotor angular speeds and, $U_1, U_2, U_3,$ and $U_4$ are the control inputs.

The quadrotor parameters are obtained based on the real parameters in [21] presented in the Tab. 1.

## III. CONTROLLER DESIGN

The main objective of the control system is to provide smooth and stable flight performance to the UAV irrespective of weather conditions. The control system is designed to perform angular stabilization, automatic control of angular position of the quadrotor, trajectory tracking, and control of the quadrotor during all flight phases from takeoff to landing.

TABLE I. QUADROTOR PARAMETERS

| Symbol and unit | Description | Value |
|---|---|---|
| m (Kg) | Quadrotor mass | 0.486 |
| l (m) | Distance between quadrotor centre of mass and propeller rotation axis | 0.25 |
| Ix (N.m/rad/s2) | Moment of inertia with respect to axe x | 3.82e-3 |
| Iy (N.m/rad/s2) | Moment of inertia with respect to axe y | 3.82e-3 |
| Iz (N.m/rad/s2) | Moment of inertia with respect to axe z | 7.65e-3 |
| d (N.m/rad/s) | Drag co-efficient | 3.23e-7 |
| b (N.m/rad/s) | Thrust co-efficient | 2.98e-5 |
| Jr (N.m/rad/s2) | Rotor inertia | 2.83e-5 |
| g (m/s2) | Earth gravity | 9.8 |

In this section, we focus only on the altitude stabilization for a given trajectory. The altitude controller keeps the distance of the quadrotor to the ground at a desired value taking into consideration that the quadrotor weight is not exactly known or changing in time. In order to achieve that an adaptive control law is proposed.

The altitude dynamics of the quadrotor can be modeled by the following equation:

$$\ddot{z} = \frac{1}{m}(\cos\phi \cos\theta)U_1 - g \quad (9)$$

The control objective is to design a control law $U_1$ So that the altitude $z$ can track a given reference $z_d$ even in the presence of uncertainty in the quadrotor dynamics, external disturbances, and sensor noise.

The sliding mode surface $S_z$ is defined as follows:

$$S_z = \dot{e}_z + \lambda_z e_z \quad (10)$$

where $\lambda_z$ is a Hurwitz positive constant, and the tracking error defined as:

$$e_z = z - z_d, \dot{e}_z = \dot{z} - \dot{z}_d \quad (11)$$

Where $z_d$ is the desired altitude, $e_z$ and $\dot{e}_z$ represent the tracking error and its derivative respectively.

The objective of the controller is to enforce the sliding mode into the surface where $S_z = 0$. In order to minimize the tracking error a Lyapunov based control is used.

Let us consider the following Lyapunov candidate function:

$$V_1(S_z) = \frac{1}{2}S_z^2 \quad (12)$$

Its first derivative is

$$\dot{V}_1(S_z) = S_z \dot{S}_z \quad (13)$$

If $\dot{V}_1(S_z) \leq 0$, then, Lyapunov asymptotic stability will be guaranteed.

Let's choose the surface $S_z$ so that $\dot{S}_z = -KS_z$, $K$ is positive constant, which implies

$$\dot{V}_1(S_z) = S_z \dot{S}_z = -KS_z^2 \quad (14)$$

Then the asymptotic stability condition is satisfied

$$\dot{V}_1(S_z) = -KS_z^2 \leq 0$$

Using equations (9), (10) and (11) the derivative of the sliding mode surface can be written as:

$$\dot{S}_z = \frac{\cos\theta \cos\phi}{m} U_1 - g - \ddot{z}_d + \lambda_z \dot{e}_z \quad (15)$$

To guarantee the stability, the condition $\dot{S}_z = -KS_z$ implies

$$U_1 = \frac{m}{\cos\theta \cos\phi}(\ddot{z}_d - \lambda_z \dot{e}_z - KS_z + g) \quad (16)$$

Since the parameter $m$ is supposed unknown, it is replaced by its estimates $\hat{m}$ in the control law

$$U_1 = \frac{\hat{m}}{\cos\theta \cos\phi}(\ddot{z}_d - \lambda_z \dot{e}_z - KS_z + g) \quad (17)$$

The closed loop dynamics become:

$$m\dot{S}_z + KS_z = -\tilde{m}(\ddot{z}_d - \lambda_z \dot{e}_z + g) \quad (18)$$

Where: $\tilde{m} = m - \hat{m}$

Equation (18) shows that, if the estimation error converges to zero then the tracking error dynamics tends to zero exponentially.
On-line estimation of the unknown parameter $m$ is related to the control law. Therefore, a Lyapunov based adaptation mechanism is applied to guarantee the asymptotic convergence.
Let the Lyapunov function candidate be given by

$$V_2(S_z, \tilde{m}) = \frac{1}{2}mS_z^2 + \frac{1}{2\gamma_z}\tilde{m}^2 \quad (19)$$

Where $\gamma_z$ is an adaptation gain. The idea is to choose the adaptation law of $\hat{m}$ such that

$$\dot{V}_2(S_z, \tilde{m}) \leq 0$$

$$\dot{V}_2(S_z, \tilde{m}) = mS_z\dot{S}_z + \frac{1}{\gamma_z}\tilde{m}\dot{\tilde{m}}$$
$$= -S_z[(\ddot{z}_d - \lambda_z \dot{e}_z + g)\tilde{m} - KS_z] - \frac{\tilde{m}\dot{\hat{m}}}{\gamma_z}$$

If the adaptation law is defined by

$$\dot{\hat{m}} = -\gamma_z S_z(\ddot{z}_d - \lambda_z \dot{e}_z + g) \quad (20)$$

Then

$$\dot{V}_2(S_z, \tilde{m}) = -KS_z^2 \leq 0$$

All the rest control and adaptation laws are obtained by following the same steps shown above.

## IV. OBSERVER DESIGN

Second-order sliding-mode super-twisting observer is proposed, in this paper, to ensure that the observer error converges to zero in finite time. Since the separation principle theorem is proved in [25], the observer can be designed separately from the controller. To ensure the separation principle for finite time convergent observers, designers prefer to switch on the controller only when the estimation error converges to zero [26].
The proposed super-twisting observer has the form

$$\begin{cases} \dot{\hat{x}}_1 = \hat{x}_2 + v_1 \\ \dot{\hat{x}}_2 = f(t, x_1, \hat{x}_2, U) + v_2 \end{cases} \quad (21)$$

Where $\hat{x}_1$ and $\hat{x}_2$ are the state estimates, and $v_1, v_2$ are the correction variables or the output error injections:

$$\begin{cases} v_1 = \alpha |x_1 - \hat{x}_1|^{\frac{1}{2}} sgn(x_1 - \hat{x}_1) \\ v_2 = \beta\, sgn(x_1 - \hat{x}_1) \end{cases} \quad (22)$$

Taking $\tilde{x}_1 = x_1 - \hat{x}_1$ and $\tilde{x}_2 = x_2 - \hat{x}_2$ we obtain the error equations:

$$\begin{cases} \dot{\tilde{x}}_1 = \tilde{x}_2 - \alpha |\tilde{x}_1|^{\frac{1}{2}} sgn(\tilde{x}_1) \\ \dot{\tilde{x}}_2 = F(t, x_1, x_2, \hat{x}_2) - \beta\, sgn(\tilde{x}_1) \end{cases}$$

And

$$F(t, x_1, x_2, \hat{x}_2) = f(t, x_1, x_2, U) - f(t, x_1, \hat{x}_2, U) + \xi(t, x_1, x_2)$$

Suppose that states are bounded, then the existence of a constant $f^+$ insured such that the inequality

$$|F(t, x_1, x_2, \hat{x}_2)| < f^+$$

Holds for any possible $t, x_1, x_2$ and $|x_2| \leq \sup|x_2|$
The observer design parameters α and β in Eq. (22), could be selected according to partial knowledge of system dynamics. Convergence of the observer states $\hat{x}_1; \hat{x}_2$ to the system state variables $x_1, x_2$ occurs in finite time, according to the theorem in [27].
Defining the flowing states variables, $x_1, x_2$ where $x_1 = z$ and $x_2 = \dot{z}$, $x_1$ represent the altitude and $x_2$ the altitude velocity. Altitude dynamics can be described by

$$\begin{cases} \dot{x}_1 = x_2 \\ \dot{x}_2 = \frac{\cos\phi \cos\theta}{m} U_1 - g + \xi(t, x_1, x_2) \end{cases}$$

The proposed super-twisting observer has the form:

$$\begin{cases} \dot{\hat{x}}_1 = \hat{x}_2 + \alpha |x_1 - \hat{x}_1|^{\frac{1}{2}} sgn(x_1 - \hat{x}_1) \\ \dot{\hat{x}}_2 = \frac{\cos\phi \cos\theta}{m} U_1 - g + \beta\, sgn(x_1 - \hat{x}_1) \end{cases} \quad (23)$$

Super-twisting sliding mode observers for the rest states are implemented in the same way as Eq. (23).

## V. SIMULATION RESULTS

In this section, we will present the simulations results of the adaptive SMC with Super-Twisting SMO applied to the quadrotor altitude control, in the presence of mass uncertainty, sensor noise and external disturbances.

Fig. 2, 3 and 4 show quadrotor states tracking a given desired trajectory, and the control effort generated using the proposed approach.

Controllers signal is converted or decoupled to be a rotor speed reference, $\omega_1, \omega_2, \omega_3, \omega_4$ as presented in Fig. 5 where the control disturbance is well rejected

Fig. 6 and 7 show the slide surface where the tracking error dynamic vanished over time and converges to zero.

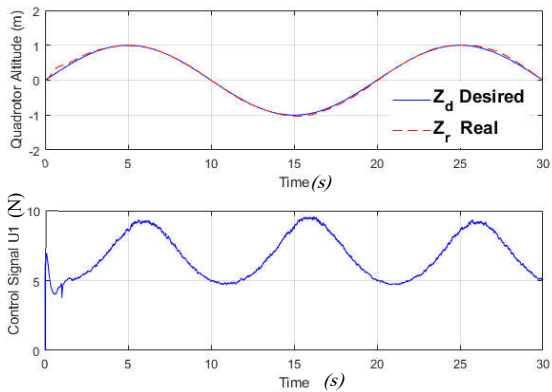

Fig. 2. Altitude Tracking and its control signal

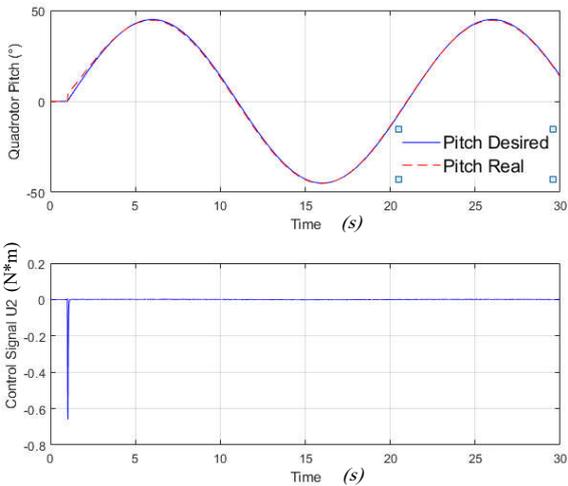

Fig. 3. Pitch tracking and its control signal

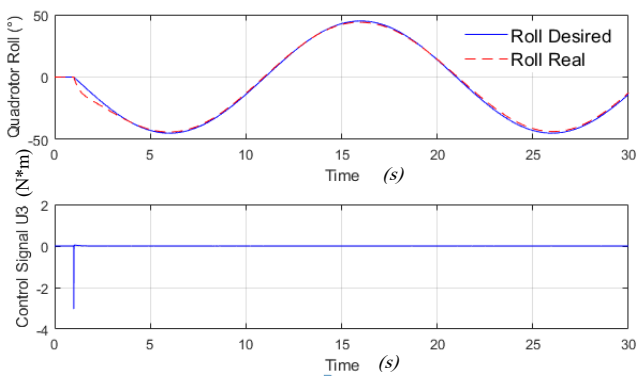

Fig. 4. Roll tracking and its control signal

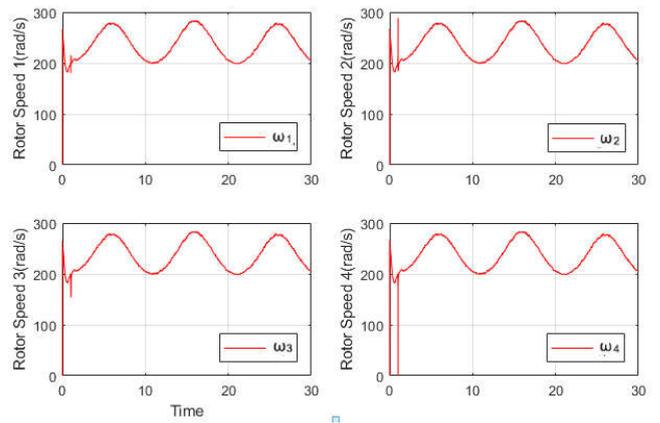

Fig. 5. The four rotors speed

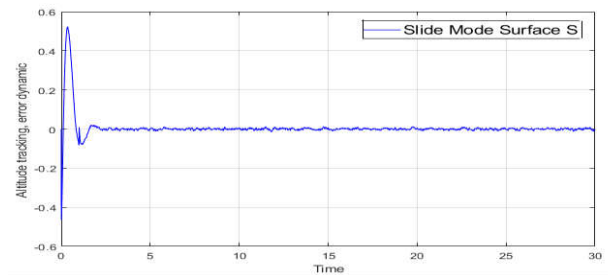

Fig. 6. *The slide surface for altitude tracking (error dynamic)*

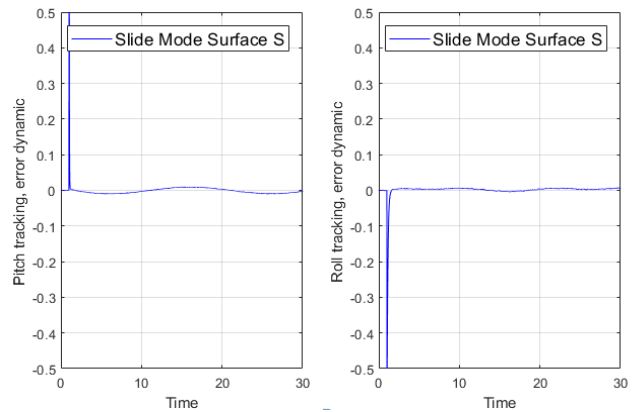

Fig. 7. The slide surface for pitch and roll tracking (error dynamic)

Fig. 8 shows the good convergence of the mass estimation to quadrotor mass, even under changes of the mass, as presented in Fig. 9 where the mass variation is tracked by the estimator and the observation error converges to zero.

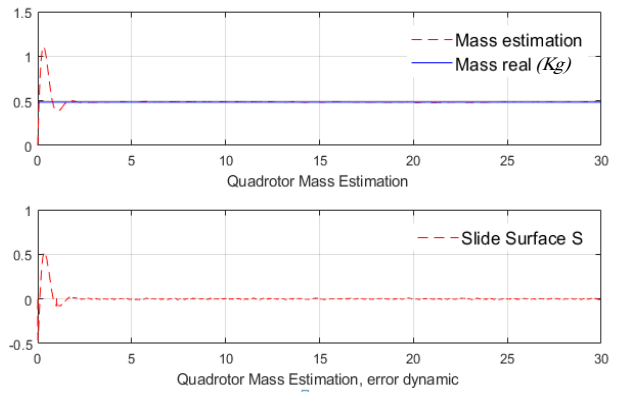

Fig. 8. Mass estimation and its slide surface

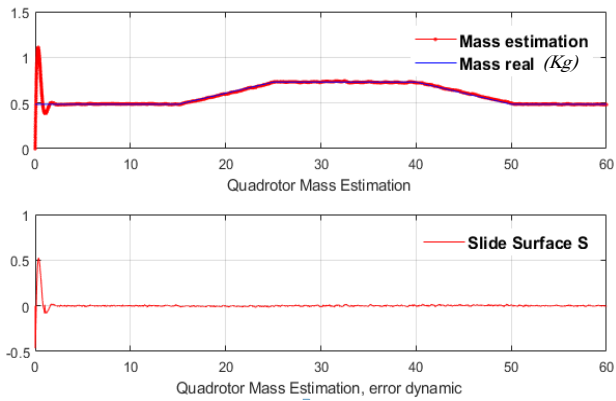

Fig. 9. Mass varying estimation and its slide surface

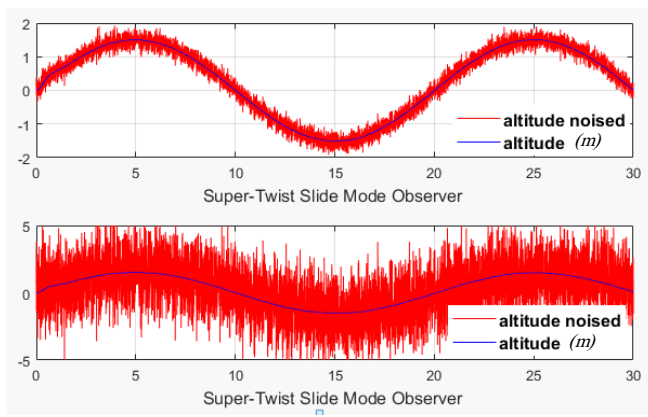

Fig. 10. Altitude slide mode observer under two different power noise

Fig. 10 shows the effectiveness of SMO even in presence of sensors noise.

## VI. Conclusion

In this paper, an adaptive sliding mode control with second-order super-twist sliding mode observer has been designed to stabilize quadrotor altitude and attitude, while tracking desired trajectories even in presence of disturbance and sensor noise.

To handle the problem of mass variation sliding mode observer has been used to estimate the quadrotor mass with finite time convergence.

Numerical simulations show the robustness and efficiency of the proposed strategy. Forthcoming works deal with the experimental implementation of the proposed approach.